\documentclass[conference]{IEEEtran}
\IEEEoverridecommandlockouts

\usepackage{cite}
\usepackage{amsmath,amssymb,amsfonts}
\usepackage{algorithmic}
\usepackage{graphicx}
\usepackage{textcomp}
\usepackage{xcolor}
\usepackage{booktabs}
\usepackage{multirow}
\usepackage{url}
\def\BibTeX{{\rm B\kern-.05em{\sc i\kern-.025em b}\kern-.08em
    T\kern-.1667em\lower.7ex\hbox{E}\kern-.125emX}}
\begin{document}

\title{Modeling Music as a Time-Frequency Image: A 2D Tokenizer for Music Generation}

\author{
\IEEEauthorblockN{
1\textsuperscript{st} Yuqing Cheng\textsuperscript{1,*,\textdagger},
2\textsuperscript{nd} Xingyu Ma\textsuperscript{2,*},
3\textsuperscript{rd} Guochen Yu\textsuperscript{2},
4\textsuperscript{th} Xiaotao Gu\textsuperscript{2}
}
\IEEEauthorblockA{
\textsuperscript{1}\textit{Department of Music AI and Information Technology, Central Conservatory of Music}, Beijing, China \\
\textsuperscript{2}\textit{Zhipu AI}, Beijing, China \\
chengyuqing@mail.ccom.edu.cn, fujindemi@gmail.com
}

\thanks{
\textsuperscript{*}Equal contribution. 
\textsuperscript{\textdagger}Work done during an internship at Zhipu AI.
}
\thanks{\copyright~2026 IEEE. Personal use of this material is permitted.  Permission from IEEE must be obtained for all other uses, in any current or future media, including reprinting/republishing this material for advertising or promotional purposes, creating new collective works, for resale or redistribution to servers or lists, or reuse of any copyrighted component of this work in other works.}
}

\maketitle

\begin{abstract}
Autoregressive music generation depends strongly on the audio tokenizer. Existing high-fidelity codecs often use residual multi-codebook quantization, which preserves reconstruction quality but complicates language modeling after sequence flattening, as the residual hierarchy imposes strong sequential dependencies and can amplify error accumulation. We propose BandTok, a generation-oriented 2D Mel-spectrogram tokenizer that represents each frame with Mel-frequency band tokens from a single shared codebook. This design yields a physically interpretable time-frequency token grid with a more independent token structure, making it better suited for autoregressive modeling. BandTok improves reconstruction with a multi-scale PatchGAN objective and EMA codebook updates. We further introduce an autoregressive language model with 2D Rotary Position Embedding (2D RoPE) to preserve temporal and frequency-band structure during generation. Experiments show that BandTok improves over residual-codebook tokenizers and achieves strong results in a data-limited setting. The source code and generation demos for this work are publicly available.\footnote{\url{https://github.com/xiaolubuhuizhuzhou/Bandtok}}
\end{abstract}


\section{Introduction}
\label{sec:intro}

Recent advances in music generation have been driven by diffusion-based generation and autoregressive token modeling. Autoregressive approaches are attractive because they leverage the scalability of language models (LMs), but their effectiveness depends critically on the audio tokenizer that converts waveforms into discrete tokens. For generation-oriented tokenization, the tokenizer must jointly satisfy high reconstruction fidelity and LM-friendly token organization. These factors determine the acoustic upper bound, sequence predictability, and error propagation behavior of autoregressive music generation.

High-fidelity neural audio codecs~\cite{zeghidour2021soundstream,defossez2022high,kumar2023high} commonly employ Residual Vector Quantization (RVQ), where multiple codebooks progressively refine reconstruction. Although RVQ preserves fine acoustic details, its residual-layer structure complicates autoregressive modeling. When multi-codebook tokens are flattened into a sequence, the LM must predict along a residual refinement hierarchy, in which later codebooks encode increasingly fine residual corrections conditioned on earlier ones. Thus, early prediction errors can propagate to later codebooks, degrading high-level token prediction and accumulating artifacts. Existing methods, including semantic-to-acoustic pipelines~\cite{borsos2023audiolm,agostinelli2023musiclm}, delayed codebook prediction~\cite{copet2023simple}, and dual-autoregressive modeling~\cite{yang2023uniaudio}, mainly address this burden at the modeling stage while retaining the residual-codebook geometry. Improving token independence has been shown to facilitate downstream LM training through independence-promoting tokenizer objectives~\cite{lemercier2024independence}.

Spectral single-codebook codecs such as MelCap~\cite{li2025melcap} and UniSRCodec~\cite{zhang2026unisrcodec} show that Mel-spectrogram-based two-dimensional tokenization can achieve compact and high-fidelity reconstruction. However, these methods are primarily evaluated as compression systems, leaving unclear whether such spectral token geometry is suitable for autoregressive music generation. In particular, a generation-oriented tokenizer must not only reconstruct well, but also produce token sequences that are stable and predictable for generation.

\begin{figure}
\centering
\includegraphics[width=0.95\linewidth]{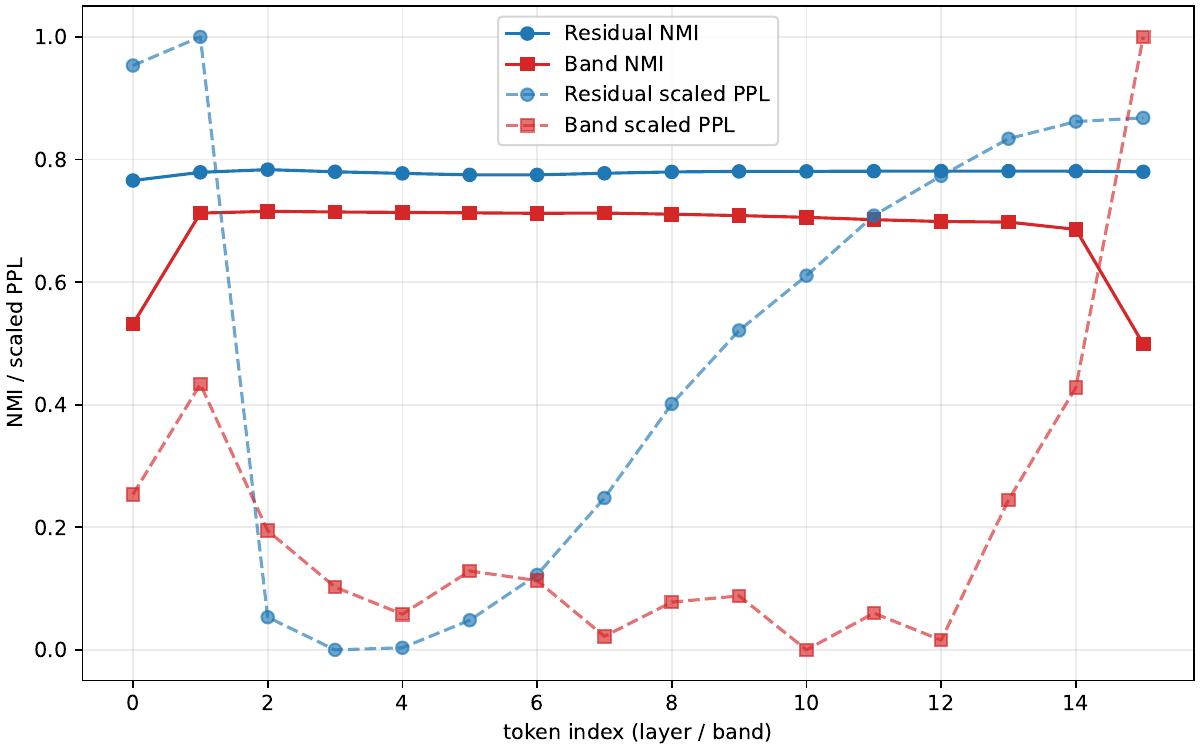}
\caption{Comparison between residual and band-wise tokens. Normalized mutual information (NMI) and language-model perplexity (PPL) are used to analyze token dependence and autoregressive prediction difficulty, respectively. Compared with residual tokens, band-wise tokens exhibit lower inter-token dependence and a more balanced PPL profile.}
\label{fig_ppl}
\end{figure}

We propose BandTok, a two-dimensional Mel-spectrogram tokenizer for autoregressive music generation. BandTok represents each frame with low-to-high Mel-frequency band tokens using a shared codebook. After flattening the time-frequency grid for LM training, the within-frame order follows spectral bands rather than residual refinement layers. Unlike RVQ, later tokens do not explicitly encode residual corrections conditioned on earlier codebooks, which reduces residual-hierarchy dependence and yields more stable autoregressive targets, as supported by Figure~\ref{fig_ppl}. We further use two-dimensional Rotary Position Embedding (2D RoPE) to preserve temporal and frequency-band position information after flattening.

To improve reconstruction fidelity and training stability, BandTok adopts a MelCap-style architecture with a multi-scale PatchGAN~\cite{isola2017image} discriminator and exponential moving average (EMA) codebook updates. The discriminator encourages perceptually important spectral detail reconstruction, while EMA stabilizes large-codebook training. Together, these designs balance high-fidelity reconstruction with LM-friendly token geometry.

Our contributions are summarized as follows:

1. We develop BandTok, a generation-oriented two-dimensional Mel-spectrogram tokenizer for autoregressive music generation. By organizing tokens along Mel-frequency bands rather than residual codebook layers, BandTok provides a physically interpretable and LM-friendly token geometry that reduces residual-chain error propagation.

2. We improve the reconstruction fidelity and training stability of spectral tokenization for music. With a MelCap-style architecture, a multi-scale PatchGAN discriminator, and EMA codebook updates, BandTok enhances high-frequency details and stabilizes large-codebook training, achieving superior reconstruction quality over waveform-domain tokenizers under comparable low-bitrate settings.

3. We introduce an autoregressive music generation framework over flattened time-frequency tokens. By incorporating 2D RoPE, the LM preserves temporal and frequency-band positional structure after flattening. Experiments show that, under comparable reconstruction quality, BandTok improves objective and subjective generation quality over alternative tokenizers and achieves stronger music generation performance under academic-scale data training.

\section{Related Works}

\subsection{Autoregressive Music Generation}

Autoregressive music generation relies on audio tokenizers that convert waveforms into discrete token sequences. Hierarchical systems such as AudioLM~\cite{borsos2023audiolm} and MusicLM~\cite{agostinelli2023musiclm} decompose generation into semantic and acoustic stages, where semantic tokens capture long-range structure and acoustic tokens reconstruct waveform-level details. However, these pipelines depend heavily on pretrained semantic representations.

MusicGen~\cite{copet2023simple} simplifies this design by directly modeling EnCodec~\cite{defossez2022high} tokens with a single autoregressive Transformer and delayed codebook prediction. UniAudio~\cite{yang2023uniaudio} further uses multi-scale Transformers and separate language models for coarse and fine tokens to handle token hierarchy. These designs improve the modeling of residual multi-codebook tokens, but they still inherit the strong inter-codebook dependence induced by residual quantization. This motivates us to revisit the tokenizer geometry itself as the interface for autoregressive music generation.

\subsection{Audio Tokenization}

Neural audio tokenizers differ in both representation domain and token geometry. Waveform-domain codecs~\cite{zeghidour2021soundstream,defossez2022high,kumar2023high,gong2026moss} achieve high-fidelity reconstruction by directly encoding waveforms, typically with residual or multi-codebook quantization. However, this structure introduces a residual codebook axis for downstream LMs, making autoregressive prediction sensitive to inter-codebook dependence and error propagation.

Spectral-domain codecs~\cite{langman2024spectral,ai2024apcodec,feng2025stftcodec} instead tokenize time-frequency representations, showing that Mel- or STFT-based representations can support high-quality audio reconstruction. However, their token streams are still primarily organized as one-dimensional or residual-codebook sequences. More recent two-dimensional spectral tokenizers, including MelCap~\cite{li2025melcap} and UniSRCodec~\cite{zhang2026unisrcodec}, exploit the image-like structure of Mel spectrograms and demonstrate strong reconstruction quality. Yet these methods are mainly evaluated as codecs, leaving the role of two-dimensional token geometry in autoregressive music generation underexplored.

\section{Method}

\begin{figure}
\centering
\includegraphics[width=0.95\linewidth]{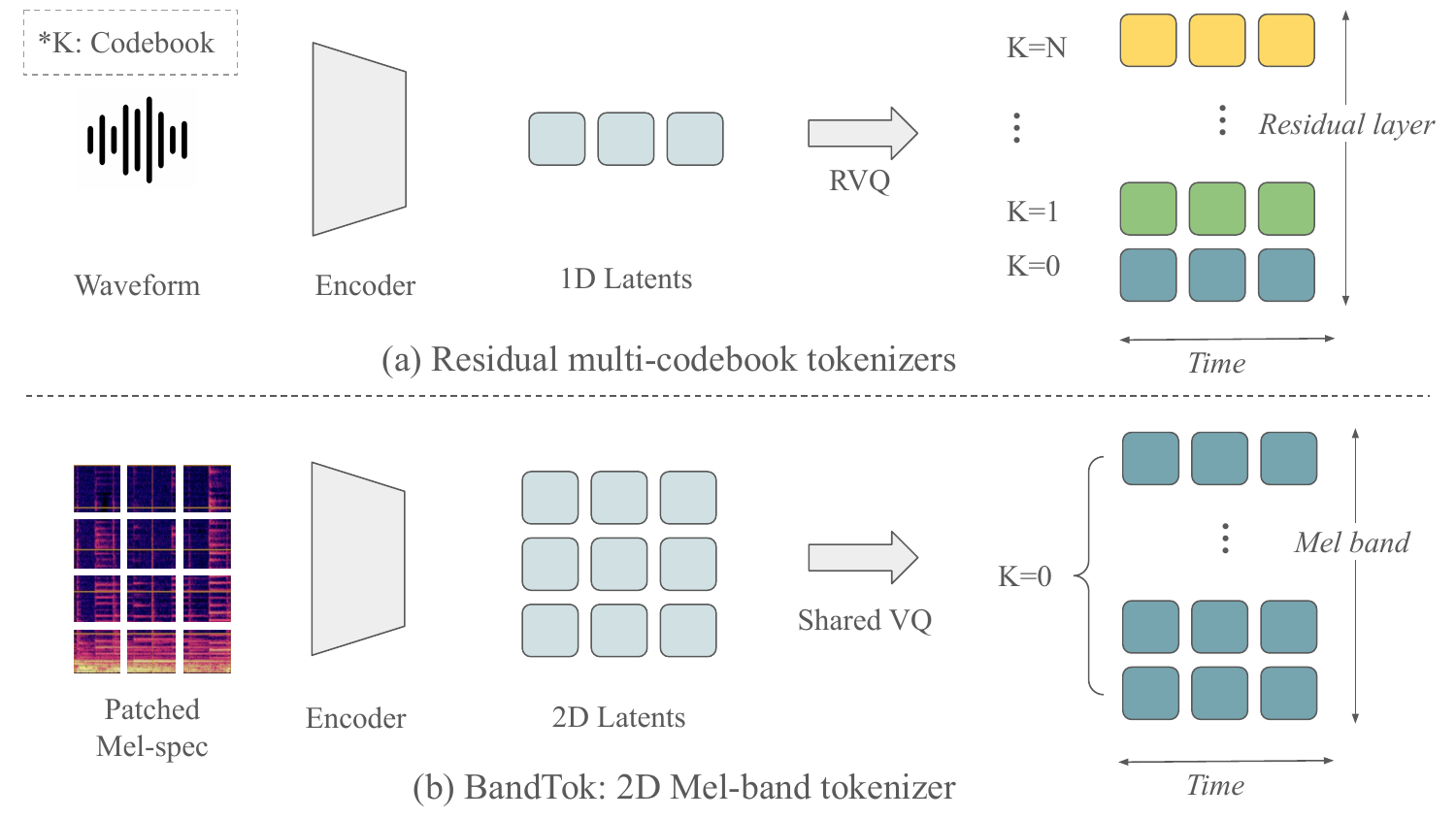}
\caption{Comparison between RVQ tokenizers and BandTok. Figure~(a) shows RVQ-based audio tokenizers, where each VQ layer quantizes the residual from the previous layer. Figure~(b) shows BandTok, which patchifies the Mel spectrogram into 2D latents and quantizes them with a single shared codebook. Its vertical axis corresponds to Mel-frequency bands.}
\label{fig_bandtok}
\end{figure}

We present BandTok, a generation-oriented Mel-spectrogram tokenizer, together with an autoregressive language model over flattened time-frequency tokens. Our method is designed around two goals: improving spectral reconstruction fidelity and preserving two-dimensional token geometry for autoregressive music generation.

\subsection{BandTok Tokenizer}

For each 44.1 kHz waveform, we compute a log-Mel spectrogram 
$\mathbf{X}\in\mathbb{R}^{N\times1\times T\times F}$ with $F=128$ Mel bins, using a 2048-sample STFT window and a 512-sample hop. BandTok first applies 2D Haar~\cite{haar1909theorie} patchification with patch size $p=2$, decomposing the spectrogram into LL, LH, HL, and HH sub-bands to retain both coarse spectral structure and local high-frequency details. A Cosmos-style~\cite{agarwal2025cosmos} encoder maps the patched spectrogram to a latent grid $\mathbf{Z}_{e}\in\mathbb{R}^{N\times C\times T'\times F'}$, downsampling by $8\times$ along both time and frequency. This yields an audio-token frame rate of approximately 10.7 Hz and $F'=16$ frequency-band positions.

As shown in Figure~\ref{fig_bandtok}, BandTok uses a single 8192-entry codebook to quantize the latent grid into a discrete two-dimensional token grid. The codebook is updated with EMA statistics instead of an explicit codebook loss, improving large-codebook stability and reducing noisy updates for rarely selected codes, while a standard commitment loss regularizes the encoder. The quantized grid is decoded back into a Mel spectrogram and converted to waveform audio using a pretrained BigVGAN-v2 vocoder~\cite{lee2022bigvgan}.

\subsection{Reconstruction Objective}

To improve high-frequency reconstruction, we introduce a multi-scale PatchGAN discriminator on Mel spectrograms. Each discriminator operates on a different spectrogram resolution obtained through linear interpolation. This design encourages realistic local time-frequency details across multiple scales, which is important for preserving musical texture and high-frequency content.

The tokenizer is trained with a weighted combination of reconstruction, perceptual, adversarial, feature-matching, and commitment losses:
\begin{equation}
\begin{aligned}
    \mathcal{L}_{\mathrm{BandTok}}
    =
    &\lambda_{\mathrm{rec}}\mathcal{L}_{\mathrm{rec}}
    +
    \lambda_{\mathrm{perc}}\mathcal{L}_{\mathrm{perc}}
    +
    \lambda_{\mathrm{adv}}\mathcal{L}_{\mathrm{adv}} \\
    &+
    \lambda_{\mathrm{fm}}\mathcal{L}_{\mathrm{fm}}
    +
    \lambda_{\mathrm{commit}}\mathcal{L}_{\mathrm{commit}}.
\end{aligned}
\end{equation}
Here, \(\mathcal{L}_{\mathrm{rec}}\) denotes the L1 Mel-spectrogram reconstruction loss, \(\mathcal{L}_{\mathrm{perc}}\) the VGG-based perceptual loss~\cite{johnson2016perceptual}, \(\mathcal{L}_{\mathrm{adv}}\) the generator-side adversarial loss, \(\mathcal{L}_{\mathrm{fm}}\) the discriminator feature-matching loss, and \(\mathcal{L}_{\mathrm{commit}}\) the VQ commitment loss. We set the corresponding weights to \(\lambda_{\mathrm{rec}}=5.0\), \(\lambda_{\mathrm{perc}}=1.0\), \(\lambda_{\mathrm{adv}}=1.0\), \(\lambda_{\mathrm{fm}}=5.0\), and \(\lambda_{\mathrm{commit}}=2.5\).

\begin{figure}
\centering
\includegraphics[width=0.95\linewidth]{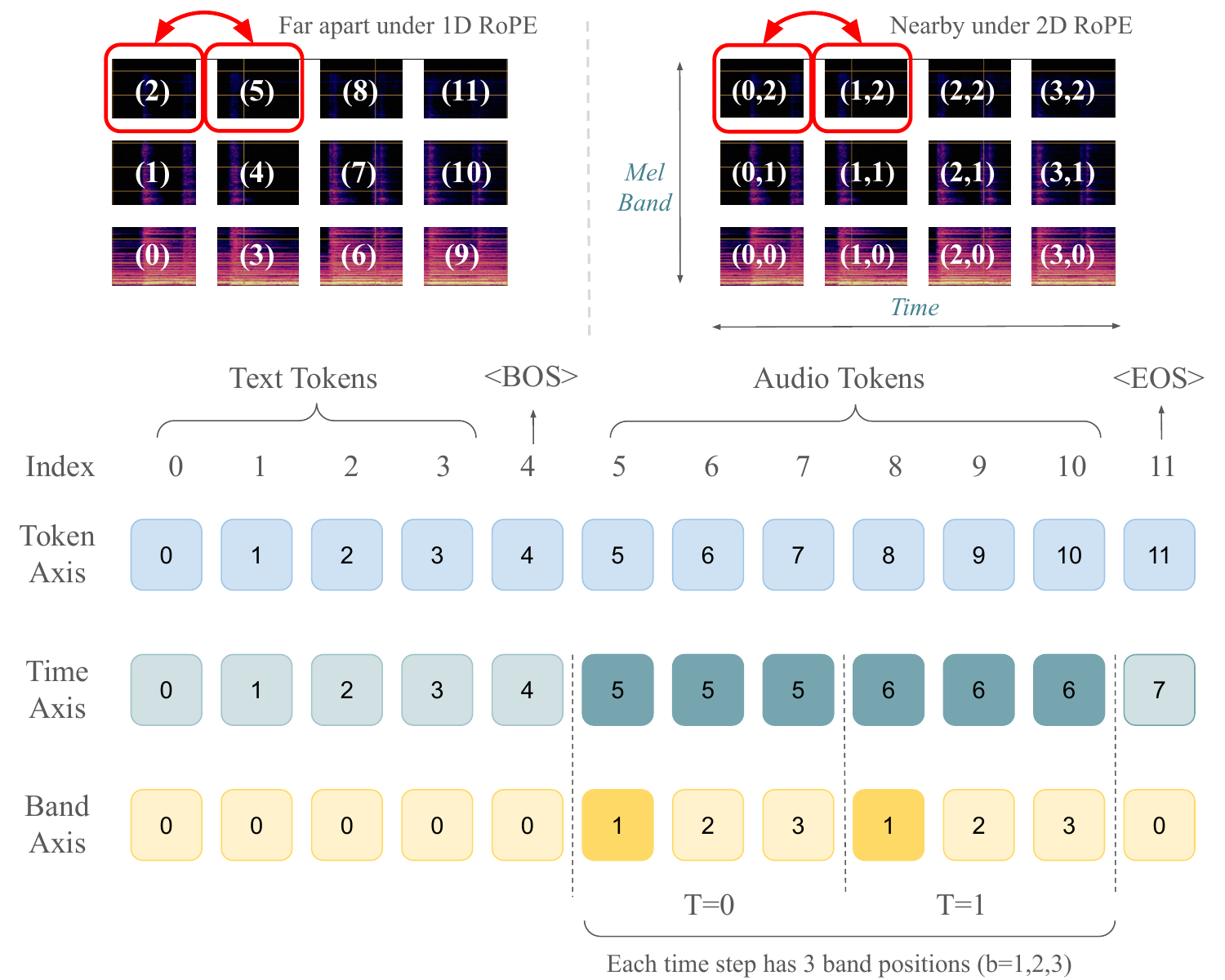}
\caption{Illustration of 2D RoPE for flattened audio tokens. It preserves the original time-frequency structure by separately encoding temporal and frequency-band positions. The token axis uses global sequence positions. The time axis follows text-token positions for text tokens and repeats each time-step index across all band tokens. The band axis is set to zero for text tokens and ranges from \(1\) to \(B\) for audio tokens within each time step.}
\label{fig_posemb}
\end{figure}

\subsection{Autoregressive Modeling with 2D RoPE}

Applying standard 1D Rotary Position Embedding (RoPE) to the flattened sequence creates a mismatch between sequence order and spectrogram geometry. In particular, tokens from the same frequency band in adjacent frames are separated by all frequency-band positions within a frame, as shown in Figure~\ref{fig_posemb}. This weakens the local time-frequency inductive bias and requires the model to infer the original two-dimensional structure implicitly.

To address this issue, we adopt Interleaved-MRoPE from Qwen3-VL~\cite{bai2025qwen3}. The attention-head dimension is split into token, time, and frequency-band components, which are interleaved at the feature level. The token axis spans the full sequence, including text, special, and audio tokens. The time and band axes explicitly encode the two-dimensional positions of audio tokens, while text tokens use zero band indices and standard sequential time indices. Under band-first flattening, all band tokens within the same frame share the same time index. This design allows the LM to retain temporal and spectral locality during autoregressive decoding.

\subsection{Conditioning}

We encode the text caption using a pretrained T5 encoder~\cite{raffel2020exploring} and prepend the resulting embeddings to the audio token sequence. Since captions describe full tracks while training is performed on shorter randomly sampled segments, we additionally encode the segment start time and the total track duration as numerical conditions. These conditions help the model distinguish, for example, an opening segment from a middle segment under the same global caption.

For classifier-free guidance (CFG), following MusicGen~\cite{copet2023simple}, we randomly replace the conditioning prefix with a near-null embedding during training. At inference time, we combine the conditional and unconditional logits as
\begin{equation}
    \ell_{\mathrm{cfg}}
    =
    \ell_{\mathrm{uncond}}
    +
    w(\ell_{\mathrm{cond}}-\ell_{\mathrm{uncond}}),
\end{equation}
where \(w\) denotes the guidance scale.

\section{Experiments}

In this section, we describe the training details and evaluate the impact of our design choices on reconstruction quality and autoregressive music generation. 

\subsection{Datasets}

For tokenizer training, we use a mixture of music and general-audio datasets, including FMA~\cite{defferrard2016fma}, Freesound~\cite{fonseca2017freesound}, MTG-Jamendo~\cite{bogdanov2019mtg}, and the MUSDB training set~\cite{rafii2019musdb18}. For language-model training, we use MTG-Jamendo with Qwen2-generated captions from the ICME 2026 Grand Challenge~\cite{hsieh2026academic}. Since we focus on instrumental music generation, we apply Mel-Band RoFormer~\cite{wang2023mel} for vocal removal and train on the resulting instrumental tracks.

For reconstruction evaluation, we randomly sample 1,000 segments from the MUSDB test set and report Mel and STFT distances. For generation evaluation, we use the official 100 contest prompts and report \(\mathrm{FAD}_{\mathrm{CLAP}}\), \(\mathrm{FAD}_{\mathrm{OpenL3}}\), and CLAP score. FAD is computed using CLAP~\cite{elizalde2023clap} and OpenL3~\cite{cramer2019look} embeddings with SongDescriber~\cite{manco2023song} as the reference dataset. We further evaluate on a 586-sample no-singing subset from SongDescriber, following Stable Audio Open~\cite{evans2025stable}, and report AudioBox~\cite{tjandra2025meta} metrics for subjective music-quality assessment, including CE, CU, PC, and PQ, corresponding to Content Enjoyment, Content Usefulness, Production Complexity, and Production Quality, respectively.

\subsection{Reconstruction Improvements}

The tokenizer is trained on 8 H800 GPUs for 24 hours with a batch size of 1024 and a segment length of 65,024 samples. We use the Adam optimizer with a learning rate of \(2\times10^{-4}\), \(\beta_1=0.8\), and \(\beta_2=0.99\). To stabilize training, we adopt an inverse learning-rate schedule with power \(0.5\), \(\texttt{inv\_gamma}=200{,}000\), and a warm-up factor of \(0.999\).

\begin{table}[t]
\centering
\caption{Ablation study of MS-PatchGAN and EMA codebook updates.}
\label{tab:reconstruction_ablation}
\begin{tabular}{lcc}
\toprule
\textbf{Model} & \textbf{Mel} $\downarrow$ & \textbf{STFT} $\downarrow$ \\
\midrule
Baseline w/ PatchGAN & 0.837 & \textbf{1.751} \\
w/ MS-PatchGAN       & \textbf{0.749} & 1.794 \\
\midrule
w/ codebook loss     & 0.763 & 1.618 \\
w/ EMA               & \textbf{0.642} & \textbf{1.544} \\
\bottomrule
\end{tabular}
\end{table}

\begin{table}[t]
\centering
\caption{Reconstruction comparison of BandTok against baseline audio tokenizers.}
\label{tab:reconstruction_overall}
\small
\begin{tabular}{lccc}
\toprule
\textbf{Model} & \textbf{Bitrate} & \textbf{Mel} $\downarrow$ & \textbf{STFT} $\downarrow$ \\
\midrule
EnCodec-32k & 2.2~kbps & 1.228 & 2.300 \\
EnCodec-48k & 3.0~kbps & 0.942 & 1.792 \\
EnCodec-48k & 6.0~kbps & 0.832 & 1.696 \\
DAC          & 2.6~kbps$^\dagger$ & 0.809 & 1.646 \\
MelCap       & 2.2~kbps & 0.730 & 1.653 \\
BandTok-1D$^\ddagger$ & 2.2~kbps & 0.690 & 1.613 \\
BandTok      & 2.2~kbps & \textbf{0.642} & \textbf{1.544} \\
\bottomrule
\end{tabular}

\vspace{2pt}
\begin{minipage}{0.95\linewidth}
\footnotesize
$^\dagger$ DAC does not provide an official 2.6~kbps checkpoint; we use the first three quantizer layers from the 8~kbps model to obtain a comparable bitrate.
$^\ddagger$ BandTok-1D denotes the RVQ variant of BandTok.
\end{minipage}
\end{table}

We ablate two reconstruction design choices for BandTok. The multi-scale Mel PatchGAN discriminator applies scale-specific discriminators to spectrograms at different resolutions and improves reconstruction over the standard PatchGAN baseline, as shown in Table~\ref{tab:reconstruction_ablation}. We also replace the conventional codebook loss with EMA codebook updates while retaining the commitment loss, which stabilizes updates for the single 8192-entry codebook and further improves reconstruction quality. Overall, BandTok achieves better reconstruction than waveform-domain tokenizers, as shown in Table~\ref{tab:reconstruction_overall}.

\subsection{Autoregressive Generation}

\begin{table*}[t]
\centering
\caption{Comparison of generation performance across different stage-I tokenizers and stage-II generators on the ICME contest test set.}
\label{tab:generation_main}
\small
\resizebox{\textwidth}{!}{
\begin{tabular}{llccccc}
\toprule
\textbf{Stage II} & \textbf{Stage I} & \textbf{Params} & \textbf{Train Data} &
\(\mathbf{FAD}_{\mathbf{OpenL3}}\) $\downarrow$ &
\(\mathbf{FAD}_{\mathbf{CLAP}}\) $\downarrow$ &
\textbf{CLAP} $\uparrow$ \\
& & & \textbf{(hours)} & & & \\
\midrule
Stable Audio Open & Stable Audio Open VAE & 1.1B & 7.3k & -- & 0.574 & 0.321 \\
MusicGen-small    & EnCodec-32k           & 300M & 20k  & -- & 0.574 & 0.370 \\
MusicGen-medium   & EnCodec-32k           & 1.5B & 20k  & -- & 0.548 & 0.353 \\
MusicGen-large    & EnCodec-32k           & 3.3B & 20k  & -- & 0.553 & \textbf{0.379} \\
\midrule
\multirow{4}{*}{Ours}
     & EnCodec-32k & 315M & 0.46k & 221.327 & 0.739 & 0.199 \\
     & EnCodec-48k & 315M & 0.46k & 266.994 & 0.898 & 0.138 \\
     & BandTok     & 315M & 0.46k & 163.804 & \textbf{0.482} & 0.163 \\
     & BandTok     & 1.5B & 0.46k & \textbf{140.006} & 0.500 & 0.171 \\
\bottomrule
\end{tabular}
}
\end{table*}

\begin{table}[t]
\centering
\caption{Generation comparison with baseline models on the Song Describer dataset.}
\label{tab:songdescriber_eval}
\small
\begin{tabular}{lccccc}
\toprule
\textbf{Stage II} & \textbf{Params} & \textbf{CE} $\uparrow$ & \textbf{CU} $\uparrow$ & \textbf{PC} $\uparrow$ & \textbf{PQ} $\uparrow$ \\
\midrule
Stable Audio Open & 1.1B & 6.725 & 7.634 & 4.342 & 7.669 \\
MusicGen-large    & 3.3B & 6.785 & 7.626 & \textbf{4.893} & 7.498 \\
\midrule
Ours              & 315M & 6.808 & 7.627 & 4.277 & 7.705 \\
                  & 1.5B & \textbf{7.244} & \textbf{7.858} & 4.040 & \textbf{7.846} \\
\bottomrule
\end{tabular}
\end{table}

We next evaluate the effect of BandTok on autoregressive language modeling. We focus on two questions: whether two-dimensional Mel-band tokens improve LM modelability, and whether 2D RoPE further improves modeling over flattened time-frequency token sequences.

To isolate the effect of token geometry, we compare BandTok with a variant denoted BandTok-1D, which uses the same model architecture but replaces the vertical Mel-band axis with residual hierarchical codebook layers. For both tokenizations, we train a 315M-parameter language model on 8 H800 GPUs for 19 hours, using a batch size of 128 and 10-second training segments. We use AdamW with a learning rate of \(5\times10^{-5}\), \(\beta_1=0.9\), and \(\beta_2=0.95\). We adopt an inverse learning-rate schedule with \(\texttt{inv\_gamma}=1{,}000{,}000\), power \(0.5\), and a warm-up factor of \(0.999\).

\begin{table}[t]
\centering
\caption{Ablation study of token geometry and positional encoding for autoregressive generation.}
\label{tab:rope_ablation}
\small
\begin{tabular}{lccc}
\toprule
\textbf{Model} & \textbf{RoPE} & \(\mathbf{FAD}_{\mathbf{CLAP}}\) $\downarrow$ & \textbf{CLAP} $\uparrow$ \\
\midrule
BandTok-1D & 1D & 1.166 & 0.117 \\
BandTok    & 1D & 0.645 & 0.193 \\
BandTok    & 2D & \textbf{0.595} & \textbf{0.214} \\
\bottomrule
\end{tabular}
\end{table}

As shown in Table~\ref{tab:rope_ablation}, BandTok improves CLAP and \(\mathrm{FAD}_{\mathrm{CLAP}}\) over BandTok-1D, indicating that Mel-band tokens are more LM-friendly than residual codebook tokens. We also compare 1D and 2D RoPE for flattened time-frequency sequences. By explicitly encoding temporal and frequency-band positions, 2D RoPE helps the LM preserve the underlying 2D structure and further improves generation quality.

\subsection{Segment-Time Conditioning}

\begin{table}[t]
\centering
\caption{Ablation study of CFG and segment-time conditioning (seg-time cond) for different LM scales.}
\label{tab:generation_ablation}
\begin{tabular}{llcc}
\toprule
\textbf{Params} & \textbf{Setting} & \(\mathbf{FAD_{\mathrm{CLAP}}}\) $\downarrow$ & \textbf{CLAP} $\uparrow$ \\
\midrule
315M & CFG \(=1.0\) & 0.700 & 0.148 \\
315M & CFG \(=2.0\) & 0.560 & 0.186 \\
315M & + seg-time cond & \textbf{0.509} & \textbf{0.206} \\
\midrule
1.5B & CFG \(=2.0\) & \textbf{0.480} & 0.217 \\
1.5B & + seg-time cond & 0.486 & \textbf{0.237} \\
\bottomrule
\end{tabular}
\end{table}

Because captions describe full tracks while training uses randomly cropped segments, we add segment-time conditioning, which encodes the segment start time and total track duration following prior work on long-form audio generation~\cite{evans2024fast}. As shown in Table~\ref{tab:generation_ablation}, this conditioning improves FAD and CLAP for the 315M model but slightly degrades FAD for the 1.5B model. We hypothesize that larger models rely more strongly on conditions and are therefore more sensitive to mismatches between fixed segment-time settings and varying track structures. Accordingly, we submit models with and without segment-time conditioning as separate variants.

Classifier-free guidance further improves generation quality. Increasing the CFG scale from 1.0 to 2.0 reduces \(\mathrm{FAD}_{\mathrm{CLAP}}\) from 0.700 to 0.560 and improves CLAP from 0.148 to 0.186.

\subsection{Token Decoupling Analysis}

We analyze whether band-wise tokenization yields a more statistically decoupled token organization. We use normalized mutual information (NMI) as a proxy for pairwise token dependence,
\[
\mathrm{NMI}(Z_i, Z_j)=\frac{I(Z_i;Z_j)}{\sqrt{H(Z_i)H(Z_j)}} ,
\]
where lower off-diagonal values indicate weaker statistical coupling across token axes.

We further evaluate autoregressive predictability using a 315M LM under a flattened token modeling scheme. BandTok-1D tokens are flattened along the residual-layer axis, whereas BandTok tokens are flattened along the frequency-band axis. We compute teacher-forced perplexity (PPL) and normalize the per-layer or per-band PPL values to $[0,1]$.

As shown in Figure~\ref{fig_ppl}, residual tokens exhibit stronger coupling and increasing prediction difficulty in later layers. In contrast, although band-wise tokens show a local PPL peak in high-frequency bands, likely due to sparse high-frequency content, they achieve lower inter-token NMI and a more balanced PPL profile across most bands. These results suggest that band-wise tokenization reduces the burden of modeling a residual hierarchy during autoregressive decoding. Both NMI and PPL analyses are conducted on the SongDescriber dataset.

\subsection{Comparison with EnCodec}

We further compare BandTok with EnCodec-32k, the waveform tokenizer used in MusicGen. EnCodec-32k represents 32 kHz audio using four 2048-entry codebooks at 50 Hz, yielding 200 tokens per second, comparable to BandTok. Under the same downstream LM architecture, BandTok achieves better generation performance, as shown in Table~\ref{tab:generation_main}.

To examine the potential effect of tokenizer pretraining data, we additionally compare against EnCodec-48k, whose reported training set includes MTG-Jamendo, while EnCodec-32k does not publicly disclose its tokenizer training data. EnCodec-48k represents 48 kHz audio using two 1024-entry codebooks at 150 Hz, yielding 300 tokens per second. As shown in Table~\ref{tab:generation_main}, EnCodec-48k performs worse than EnCodec-32k, likely because its higher token rate and larger token space increase the downstream modeling burden.

\subsection{Scaling the Language Model}

We study LM scaling by comparing 315M and 1.5B models trained on the same data. As shown in Table~\ref{tab:generation_main}, on the primary evaluation set, increasing model size does not consistently improve $\mathrm{FAD}_{\mathrm{CLAP}}$, suggesting that larger LMs may require more training data and more diverse captions. Nevertheless, the 1.5B model improves instrumental timbre recognizability, as reflected by better $\mathrm{FAD}_{\mathrm{OpenL3}}$ and CLAP score.

To further assess generation quality, we evaluate on a larger SongDescriber subset. As shown in Table~\ref{tab:songdescriber_eval}, our model achieves the best AudioBox scores among the baselines on this larger set, demonstrating strong generation quality under an academic-scale training setup. While CLAP score remain limited, likely due to the prefix-based text-conditioning strategy, these results highlight the potential of modeling music with two-dimensional time-frequency tokens.

\section{Conclusion}

We presented \emph{BandTok}, a generation-oriented 2D Mel-spectrogram tokenizer for autoregressive music generation. By replacing residual codebook layers with physically meaningful Mel-frequency band tokens, BandTok provides an LM-friendly time-frequency token geometry while maintaining high reconstruction fidelity. With 2D RoPE, BandTok preserves temporal and spectral structure during decoding and improves generation quality over residual-codebook tokenizer baselines. Future work will improve text following through better condition control and caption augmentation, and extend this paradigm to broader audio generation tasks.

\bibliographystyle{IEEEbib}
\bibliography{icme2026references}

\end{document}